# Plasmonic Heating in Au Nanowires at Low Temperatures: The Role of Thermal Boundary Resistance

Pavlo Zolotavin[1], Alessandro Alabastri[1], Peter Nordlander[1,2,3], Douglas Natelson[*,1,2,3]

[1]Department of Physics and Astronomy, Rice University, 6100 Main St., Houston, Texas 77005, United States

[2]Department of Electrical and Computer Engineering, Rice University, 6100 Main St., Houston, Texas 77005, United States

[3]Department of Materials Science and NanoEngineering, Rice University, 6100 Main St., Houston, Texas 77005, United States

[*]E-mail: natelson@rice.edu.

**ABSTRACT**

Inelastic electron tunneling and surface-enhanced optical spectroscopies at the molecular scale require cryogenic local temperatures even under illumination – conditions that are challenging to achieve with plasmonically resonant metallic nanostructures. We report a detailed study of the laser heating of plasmonically active nanowires at substrate temperatures from 5 to 60 K. The increase of the local temperature of the nanowire is quantified by a bolometric approach and could be as large as 100 K for a substrate temperature of 5 K and typical values of laser intensity. We also demonstrate that a ~3× reduction of the local temperature increase is possible by switching to a sapphire or quartz substrate. Finite element modeling of the heat dissipation reveals that the local temperature increase of the nanowire at temperatures below ~50 K is determined largely by the thermal boundary resistance of the metal–substrate interface. The model reproduces the striking experimental trend that in this regime the temperature of the nanowire varies nonlinearly with the incident optical power. The thermal boundary resistance is demonstrated to be a major constraint on reaching low temperatures necessary to perform simultaneous inelastic electron tunneling and surface enhanced Raman spectroscopies.

Many of the benefits that can be derived from surface plasmon resonances are associated with increased light scattering and absorption at the resonance wavelength, and that absorption leads to an increase of local temperature of the metal and surrounding media. For most common applications of plasmon resonance like molecular sensing,[1,2] sub-wavelength field manipulation,[3–5] and local field enhancement using metal nanostructures,[6,7] heating is usually of little concern. Local plasmonic heating becomes significant in isolated nanostructures.[8] Plasmon-driven heating has been extensively studied at room temperature and successfully utilized for targeted thermal treatment of cancer tumors,[9] driving chemical reactions,[10] and control of the material temperature with nanoscale localization.[11–14] Local plasmon-driven temperature variation has also been proposed as an essential driver of photo-induced changes in local electrostatic potential.[15]

At low temperatures the optical properties of plasmonic nanostructures could be strongly dependent on local temperature, which could be used as an efficient control parameter.[16] On the other hand, an unaccounted increase in local temperature is detrimental for experiments that require cryogenic environments, as in the studies of simultaneous electron tunneling and surface enhanced Raman spectroscopies of single molecule conductors.[17–23] Understanding the heat dissipation, through the quantification of plasmonic heating in metal nanostructures at low temperature, is crucial for these kinds of experiments.

A variety of mostly optical methods have been developed to measure the optically driven temperature increase of plasmonic nanostructures at or above room temperature.[24–32] At low temperatures, $T \sim 80\ K$, a bolometric method was used to assess the temperature increase in gold nanowire attached to larger metal leads.[33] For a given structure and input optical power, the local heating (a few Kelvin) with 80 K substrate temperature was approximately twice as large as the local heating with the environment at room temperature. Following this trend, one would expect the local heating to be further increased when the sample environment is held at liquid helium temperatures. Several factors contribute to the larger local temperature increase and complicate the analysis of thermal dissipation in the cryogenic environment. At lower overall temperatures, the phonon thermal boundary resistance that arises due to the acoustic mismatch between interfacing different materials is significantly enhanced, preventing efficient heat transfer through the metal-substrate interface.[34] Analysis is further complicated by the strong temperature dependence of the thermal boundary resistance and its dependence on the surface quality of the interface.[35] Secondly, the thermal conductivity of most materials has a non-linear and non-monotonic temperature dependence below ~50 K. These effects can be exacerbated in some limits, such as when heater size becomes comparable to the phonon mean free path,[36,37] and when low energy phonon availability is limited.[38] Moreover, as the local metal temperature is increased and induced heating also increases due to the temperature dependence of the dielectric function of gold. These factors combine, leading to a significantly enhanced local temperature and a non-linear dependence of local temperature on the laser intensity or substrate temperature. Note that in the present work we are *not* examining the extremely high heat flux regime reached in ultrafast pulsed laser experiments. As a consequence, we consider the steady-state situation of quasi-equilibrium between the electrons and the lattice, such that it is reasonable to describe these two systems as sharing a locally defined temperature.

In this work we report studies of plasmon-induced local heating in Au nanowires fabricated on $SiO_2$/Si, sapphire, and quartz substrates, using a similar bolometric measurement technique as in Ref. 33, but extending the temperature range down to the cryogenic limit of interest for electronic spectroscopy experiments. As the substrate temperature is reduced from 60K to 5K, the magnitude of plasmonic heating increases and is substantially larger than that previously reported for higher substrate temperatures, exhibiting a strongly nonlinear dependence on the incident optical power. Finite element modeling of the plasmonic and direct optical absorption and heat transport fully reproduces experimental observations quantitatively with only one adjustable parameter, the acoustic mismatch thermal boundary resistance between the metal and the dielectric substrate. The modeling confirms that the observed non-linear dependence of heating on laser intensity results from the strong temperature dependence of the thermal boundary resistance between metal layer and the substrate. The modeling reveals that the local temperature in the center of the nanowire is larger than the volume-average temperature that is estimated from the bolometric measurement of the entire device's resistance, due to the finite resistance of the wider contact leads. Additionally, when the polarization of the laser radiation is chosen such that the plasmon mode is not excited, the local temperature of the nanowire is predicted to be lower than that inferred in the experiment, due to the fact that wider leads have larger geometrical cross section compared to the nanowire itself.

**Results and Discussion**

Fig. 1a is a scanning electron image of a typical bowtie device that consists of a nanowire contacted with wider leads that provide connection to large metal pads. The length of the nanowires in this work was 550-600 nm, with the width chosen so that the transverse plasmon resonance of the wire is excited by radiation of the 785 nm CW diode laser. The optimal nanowire width for each substrate material depends on the dielectric environment and was determined experimentally. The Au thickness was 14 nm with an additional 1 nm Ti adhesion layer and was kept the same for all devices studied. Additional details of the device fabrication can be found in the Methods section.

The heating of the nanowire under laser illumination was measured using the bolometric approach. The total electrical conductance of the bowtie structure is primarily limited by the conductance of nanowire constriction, because the width of the fan-out leads, 10 μm, is significantly larger than the width of the nanowire. A local increase in nanowire temperature results in reduction of the total conductance. This could be used to evaluate the heating if the temperature dependence of the conductance, $G(T)$, is previously recorded. At $T \geq 40\ K$, knowledge of the $dG/dT$ is sufficient to calculate the change in nanowire temperature because $G(T)$ is close to linear in $T$.[33] For low temperatures, $G(T)$ is no longer linear in $T$ and a fit to the full, empirically measured functional form $G(T)$ is necessary for an accurate $G \to T$ conversion.

The photo-induced change in conductance is defined as $\Delta G = G(T_{dark}) - G(T_{light})$, where $T_{dark}$ is the temperature of the device without laser illumination, *i.e.* the temperature of the substrate, and $T_{light}$ is a temperature of the nanowire under laser light. $\Delta G$ was measured by a lock-in amplifier, which was synced to a mechanical chopper modulating laser light intensity at 339 Hz. At this frequency thermalization occurs much faster than the period of laser intensity modulation.[39,40] The devices were biased at 5 mV DC for this measurement. We verified that DC bias up to 20 mV was small enough to produce negligible change in device temperature compared to the effects of laser heating. A second lock-in amplifier was synced to the 3 mV AC bias at 439 Hz and was measuring the total conductance of the device. When laser radiation is blocked this lock-in amplifier measures, $G(T_{dark})$ and which was recorded during cool-down. Knowledge of $G(T_{dark})$ and $\Delta G$ is sufficient to calculate the effective temperature increase, $\Delta T = T_{light} - T_{dark}$, of the nanowire due to laser illumination, provided that the total conductance of the device is determined by the conductance of the nanowire and the thermal gradient is negligible. This assumption is equivalent to saying that $G(T_{dark}) = G(T_{light})$, when $T_{dark} = T_{light}$. Limitations of this approach are revealed by numerical calculations and discussed later. Additionally, under the chopped laser illumination the second lock-in amplifier measures the average of $G(T_{dark})$ and $G(T_{light})$ because we used a lock-in time constant of 100 ms, which is considerably larger than the period of the laser intensity modulation. This measurement offers a second method to calculate the temperature increase. The $\Delta T$ calculated from these two measurements agree within 0.25 K and were carried out as an additional cross check during data analysis. In the main text we report the $\Delta T$ obtained using the first method, as it produced data with a better signal to noise ratio. Heating was observed in all 90 fabricated devices and detailed measurements for different substrate temperatures, laser powers, and polarizations were performed for 21 of these.

Results of the polarization dependent $\Delta T$ measurement for nanowires fabricated on the $SiO_2$/Si substrate are demonstrated in Fig. 1b for varying substrate temperatures.

The angular dependence of $\Delta T$ is determined by a sum of the $\cos^2(\theta)$ contribution from *transverse* plasmon resonance, with maximum close to $\theta = 90°$, and non-resonant absorption of laser radiation that determines heating at $\theta = 0°$. (Because the metal system is *essentially infinite in extent* along the $\theta = 0°$ direction, there is no longitudinal local plasmon resonance to be excited with that incident polarization.) As the substrate temperature decreases from 60 K to 5 K, the laser induced heating increases reaching almost 100 K at $\theta = 90°$. Fig. 1b demonstrates experimental results for the laser intensity of 180 kW/cm² (averaged over the beam spot), which corresponds to 4.5 mW – a typical laser power used in surface-enhanced Raman spectroscopy (see SI for data at other intensity levels).[41–43] In addition to the substantial heating, the shape of the $\Delta T(\theta)$ evolves as the substrate temperature is reduced. This can be readily seen from comparing the $\Delta T(90)/\Delta T(0)$ ratio: for $T = 5\,K$ the ratio is 1.8 and at $T = 60\,K$ the ratio is 2.3. The dependence of $\Delta T$ on laser intensity is also non-trivial and is demonstrated in Fig. 1c. At $T = 75\,K$, $\Delta T$ increases almost linearly with laser intensity. As the substrate temperature is reduced, the $\Delta T$ at low intensity levels demonstrates a sharp non-linear upturn before leveling into an approximately linear dependence with similar slopes for different substrate temperatures. When the absolute local temperature (substrate temperature $T$ plus local $\Delta T$ from the heating) of the device exceeds ~ 50 K (region where the data is linear in Fig 1c), the absolute local temperature as a function of incident intensity becomes linear (see Fig. S2). As a result of the nonlinear dependence of $\Delta T$ on intensity at low temperatures, in the high intensity limit the absolute temperature range is much more limited (extending from 130 K to 150 K) than the initial substrate temperature range (from 5 K to 75 K). We choose to plot the data in terms of the local heating amount $\Delta T$ to emphasize the physics limiting the heat transfer from the optically heated metal.

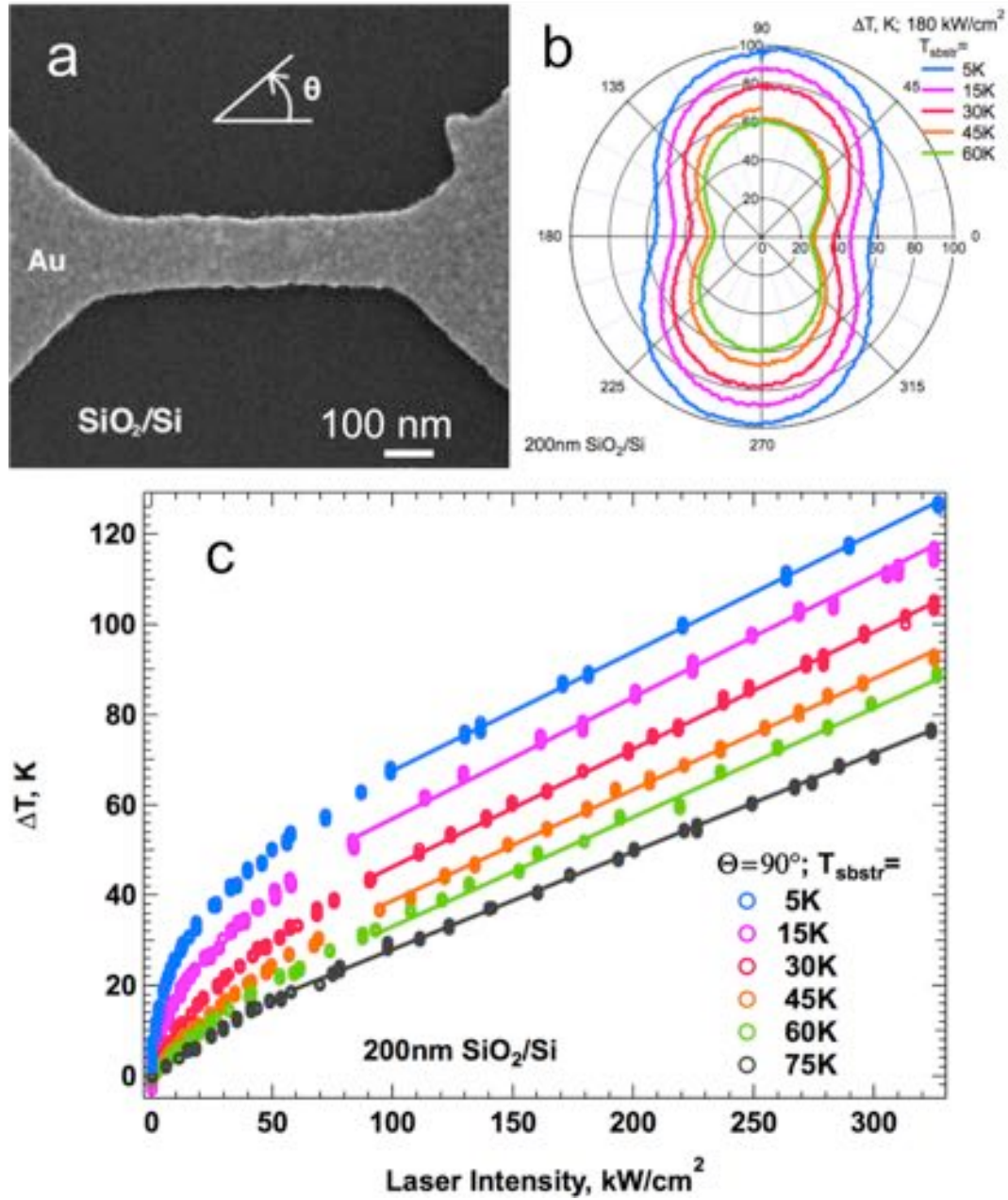

**Figure 1**. Temperature increase for bowties fabricated on the SiO$_2$/Si substrate: (a) Scanning electron microscope (SEM) image of a typical device. Width of the device is 138 ± 5 nm, length is 600 ± 30 nm, Au thickness is 14 nm plus 1 nm of Ti adhesion layer. (b) Polarization dependence of the light-driven temperature increase at different substrate temperatures for 180 kW/cm$^2$ laser intensity, with 785 nm incident wavelength and the laser focal spot diameter is 1.8 μm. (c) Dependence of the temperature increase on laser intensity. Lines through data points serve as a guide to the eye.

The steady-state temperature of the nanowire under CW laser illumination is determined by the efficiency of the thermal transfer from the nanowire to the fan-out Au electrodes and to the substrate. The thermal conductivities of the gold film, substrate, and the thermal boundary resistance, $R_{bd}$, between metal and substrate are therefore parameters that determines the final nanowire temperature. The thermal boundary resistance, well known in low temperature physics circles but not as familiar to the plasmonics community, arises due to the differences in electronic and vibrational properties between interfacing materials and increases as temperature is reduced. For solid-solid interfaces it is well approximated by an acoustic mismatch model that predicts inverse cubic temperature dependence.[34] The effects of thermal boundary resistance on heat dissipation from metal nanostructures at room temperature is very small and is not usually included in modeling.[44] At low temperatures, however, inclusion of this physics is crucial for development of accurate models.

A large local temperature approaching 100 K is undesirable for cryogenic applications, but could of course be reduced to an acceptable level by decreasing the laser intensity, at the cost of decreased signal from any optical measurement. To find which of the material properties contribute the most to the large light-driven temperature increase we performed a model numerical calculation to estimate the magnitude and trends of the observed effects. We note that temperature increase due to plasmonic excitation is large compared to the substrate temperature and occurs over the range where the thermal conductivity, $\kappa(T)$, electrical conductivity, $\sigma(T)$, and thermal boundary resistance, $R_{bd}(T)$, of the various materials all have non-linear temperature dependences. This prevents the development of a simple analytical solution, and a numerical solution of the heat dissipation problem is required.

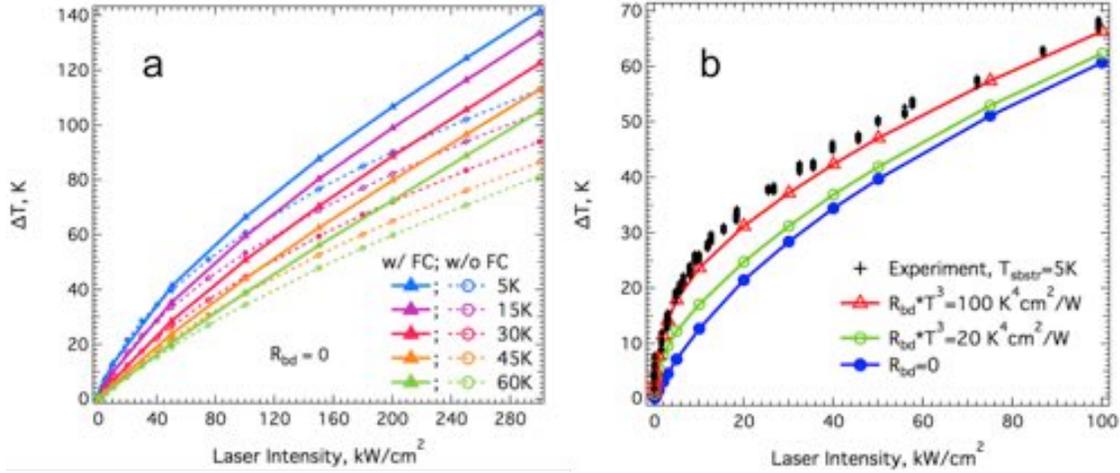

**Figure 2**. (a) Simulation of the temperature increase in the center of the nanowire using experimental parameters from Fig. 1c at different substrate (200 nm SiO$_2$ on Si) temperatures (5-60 K) without including the thermal boundary resistance. Solid lines depict result of the calculations including the full coupling (FC) between the local temperature and temperature-dependent dielectric properties of gold, and dashed lines without including the temperature dependence of the gold dielectric function (w/o FC). (b) Evaluation of the effects of the thermal boundary resistance on the temperature increase. Calculations are performed without FC at $T_{sbstr} = 5K$.

Finite-element numerical modeling was performed using the COMSOL Multiphysics, v5.1 software suite. The 3D model included an option of the full coupling of electromagnetic and heat transfer equations. This approach allowed us to take into account the temperature dependence of Au, Si, SiO$_2$, sapphire thermal conductivities and the Au dielectric function.[16,45,46] The numerical model, Fig. 2, successfully reproduces the experimental observation of the non-linear dependence of heating on laser intensity and the evolution towards linearity at large substrate temperature, for the 200 nm SiO$_2$ on Si substrate. We first consider the case without including a thermal boundary resistance. We run two versions of the calculation. In the full coupling (FC) model, we start with the dielectric function for the gold at the initial and uniform temperature, $\varepsilon_{Au}(T_{init})$. We then iterate, solving the electromagnetic and thermal transport problems together with the local $\varepsilon_{Au}(T)$ evaluated at each step and at every mesh element, until the system converges to the final steady-state temperature distribution. In the no-FC calculation, we use $\varepsilon_{Au}(T_{init})$ all the time. Fig 2a demonstrates that the model without FC underestimates the $\Delta T$ at large laser intensity as it does not take into account changes in dielectric function of gold as the metal temperature is increased. At low laser intensity, when the $\Delta T$ is small, both models predict similar local temperature, which justifies the use of the model without FC for $\Delta T$ smaller than ~50 K. Incorporation of the thermal boundary resistance in the model leads to the increased $\Delta T$; however this effect is significant only for $\Delta T < 50K$, Fig. 2b. The best agreement with the experimental data is obtained for $R_{bd}T^3 \sim 10^2\ K^4 cm^2/W$. Comparison with previously published data is difficult due to the fact that exact value of $R_{bd}(T)$ depends on the surface quality and metal deposition parameters.[34,47,48] While more sophisticated models are possible, the minimalistic numerical model without full coupling calculation, but including the thermal boundary resistance, is sufficient to reproduce the observed experimental trends.

      The simulation predicts that the peak temperature rise in the center of the wire is larger than the area-averaged temperature rise inferred from the total resistance change. This is not surprising, given that the total device resistance is not given by that of the wire alone, but includes a contribution from the larger contact regions near the wire that remain at the substrate temperature. The discrepancy between the true local temperature rise and the temperature change inferred from the total resistance would therefore be more severe in longer wire geometries, and could be minimized by optimization of device geometry

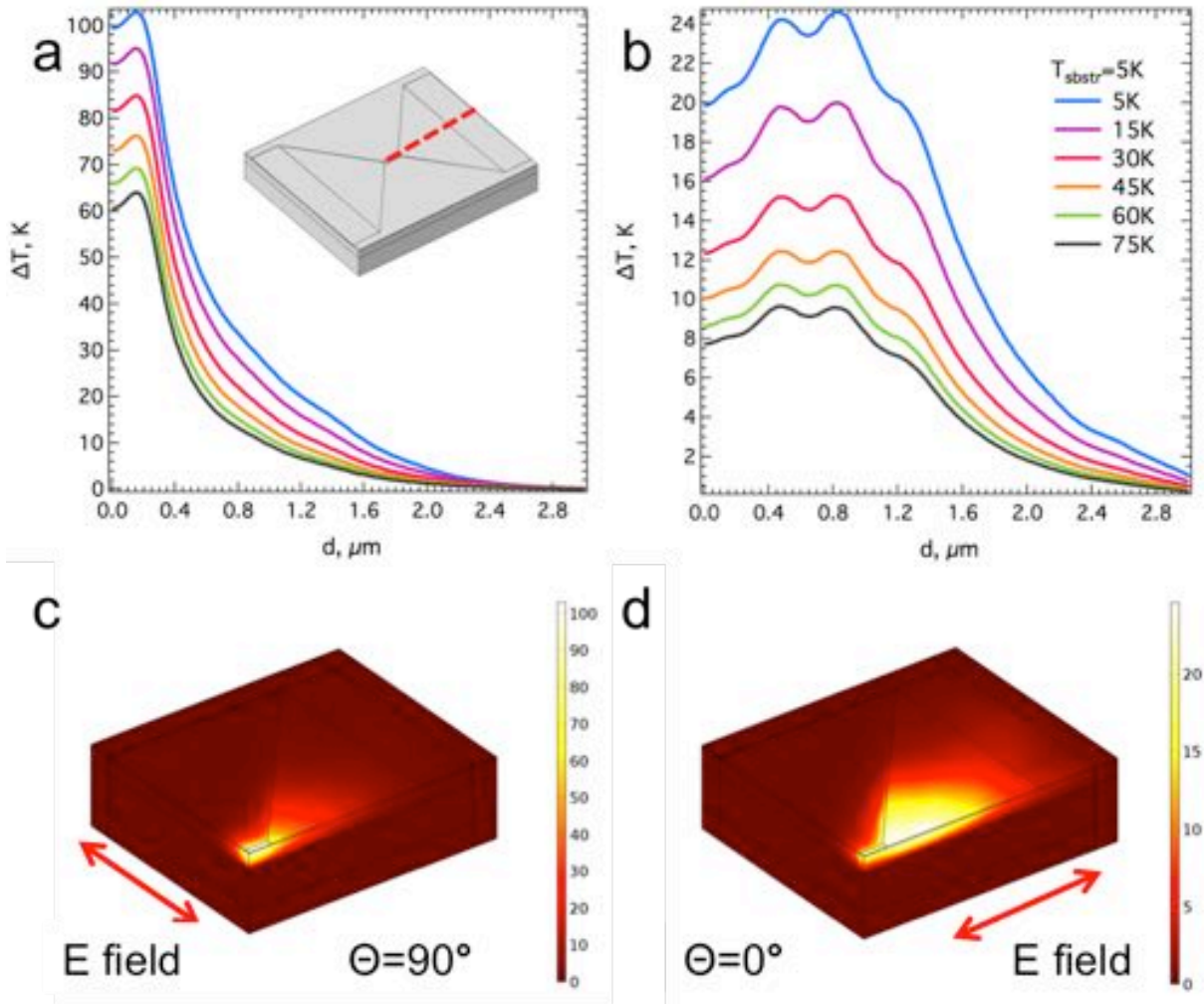

**Figure 3**. (a) Local temperature increase along the center of the device fabricated on SiO$_2$/Si substrate (red dashed line in the inset simulation geometry schematics) for different substrate temperatures (5 - 60 K) and transverse light polarization. (b) The same as in (a), but with longitudinal polarization. (c), (d) 3D $\Delta T$ distribution maps corresponding to the $T_{sbstr} = 5K$ from (a) and (b) respectively. The temperature profile inside the gold film is essentially uniform in the vertical direction. Calculations were performed using the model without full coupling as in Fig. 2a.

Fig. 3 shows the calculated temperature distribution along the center of the nanowire and in 3D for the SiO$_2$/Si substrate. In the case of the transverse plasmon mode excitation, $\theta = 90°$, the maximum temperature is located in the center of the nanowire and the overall heat distribution map resembles that of the heat dissipation from a point like heat source. The region of elevated temperature extends beyond the length of the nanowire for two reasons. Firstly, due to the presence of thermal boundary resistance and the fact that thermal conductivity of gold is larger than that of SiO$_2$, there is a lateral thermal pathway through the gold film adjacent to the nanowire to the substrate, in addition to direct heat transfer from the gold nanowire into the adjacent substrate. Secondly, the laser spot is larger than the length of the nanowire, which results in some heating due the direct absorption of laser radiation by the gold film of the electrodes. When the plasmon mode is not excited, $\theta = 0°$, the nanowire has a lower temperature than the adjacent area of the fan-out leads. Recall that the extended (effectively infinite) nature of the metal along the $\theta = 0°$ direction means that there is no local plasmon to be excited with this polarization. The more complex temperature profile occurs because in the absence of plasmon excitation the heating is now caused by direct absorption of laser radiation by gold film alone, which is determined by the geometrical overlap between the metal film and the laser spot. For this temperature distribution the bolometric measurement should overestimate the $\Delta T$ of the nanowire as the area adjacent to the nanowire has higher resistance. The peaks in $\Delta T(d)$ visible in Fig. 3a and 3b are robust features of the simulation, insensitive to changes in finite element meshing or the simulation space size, and likely result from interference between propagating plasmon modes excited from the edges of the metal (see SI for calculation results in different simulation geometries). Due to the computational complexity when considering polarization not along structural symmetry directions, we did not perform detailed calculations at other polarization angles. However, based on the two angles examined, we infer that evolution of the shape of $\Delta T(\theta)$ as the substrate temperature is increased is likely to be caused by the effects of the changing temperature distribution. This and the fact that heating of the connecting electrodes leads to a difference between the inferred $\Delta T$ and actual local nanowire temperature further highlights the need for detailed calculations of heat dissipation if a precise determination of the local temperature is a necessity.

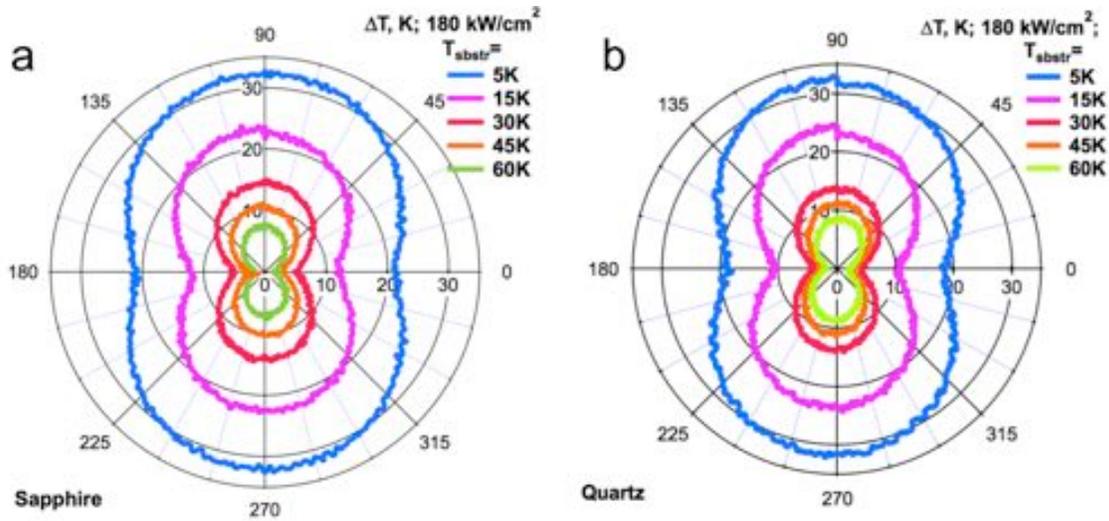

**Figure 4**. Polarization dependence of the temperature increase for nanowire devices fabricated on the sapphire, (a), and quartz, (b), substrates. Data is shown for laser intensity of 180 kW/cm². Additional data at other intensity levels is available in SI.

Simulations suggest that for nanowires fabricated on thermally oxidized Si wafers, the heat dissipation is primarily governed by the low thermal conductivity of $SiO_2$. Motivated by the desire to reduce the unwanted heating under laser illumination, we fabricated sets of devices on sapphire and quartz substrates. Switching to single-crystal substrates resulted in a threefold reduction of $\Delta T$ under the same level of laser intensity, as demonstrated in Fig. 4. We attribute small differences in $\Delta T(\theta)$ between quartz and sapphire substrate to the geometrical variations between different devices. The nanowire width on the sapphire substrate was reduced to 120 ± 5 nm to maintain the plasmon resonance at a free space incident wavelength of 785 nm. The polarization dependence of the $\Delta T$ on sapphire and quartz substrates is similar to that on the $SiO_2$/Si. Again, as the substrate temperature is reduced, the shape of the polarization dependence becomes more circular for both substrate materials.

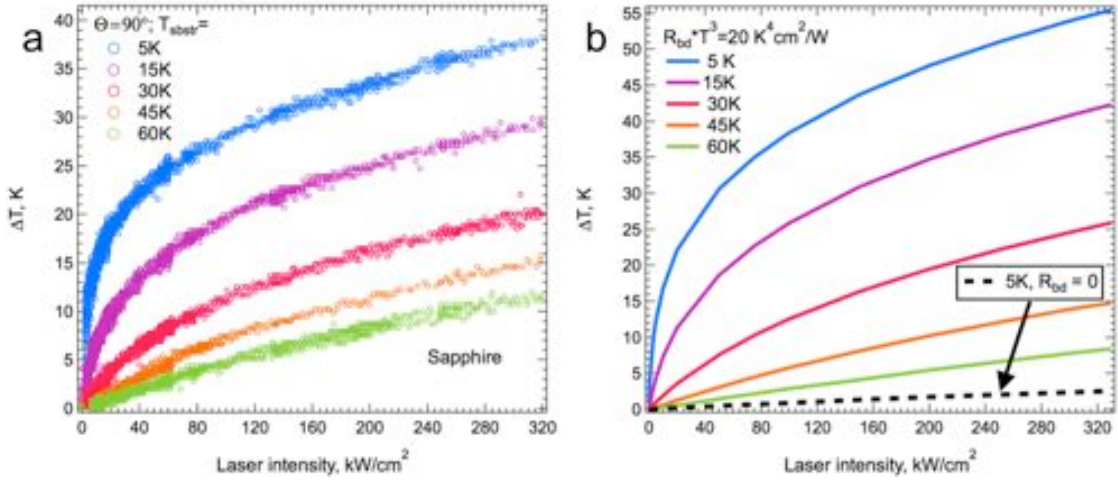

**Figure 5**. (a) Dependence of the temperature increase of the nanowire on laser intensity for the nanowire fabricated on a sapphire substrate. Laser polarization is set perpendicular to the nanowire, $\theta = 90°$. (b) Results of the numerical simulation using the same geometry as in (a) for different substrate temperatures (5 - 60 K). The case without the thermal boundary resistance, dashed line, is calculated for $T_{sbstr} = 5K$. Calculations were performed using model without full coupling.

The dependence of the $\Delta T$ on the laser intensity for sapphire substrate, Fig. 5a, displays the same qualitative features as was demonstrated in Fig. 1c for the SiO$_2$/Si substrate. At $T_{sbstr} = 5K$ the thermal conductivity of sapphire and crystalline quartz is ~10$^3$× larger than that of amorphous SiO$_2$ at this temperature; however, the $\Delta T$ is reduced only by a factor of around three. This suggests that it is the thermal boundary resistance rather than the bulk sapphire thermal conduction that plays the key role for limiting thermal transport out of devices fabricated on the sapphire and quartz. Indeed, without addition of the thermal boundary resistance the predicted heating is less than 3 K, Fig. 5b. The magnitude of the calculated temperature increase strongly depends on the chosen value for the $R_{bd}$. Calculation results demonstrated in Fig. 5b were obtained with the $R_{bd}T^3 = 20\ K^4 cm^2/W$, which is close to the theoretically predicted $19\ K^4 cm^2/W$.[35] Results of the calculations at other values of the $R_{bd}$ are demonstrated in SI. We emphasize that this acoustic mismatch boundary resistance contribution must be present due to the different acoustic properties of the metal nanowire and dielectric substrate.

We now consider possible additions to our model. Our thermal transport model assumes validity of the Fourier law throughout the whole temperature range studied. If the phonon mean free path exceeds the characteristic size of the heater, the ballistic heat transfer should be considered.[36] It was demonstrated that this effect could be approximated as a reduction of the bulk thermal conductivity and could potentially contribute to the additional heating not accounted by our model.[37] We explored this possibility by performing a set of test calculations with reduced thermal conductivity of sapphire by as much as a factor of 10 and did not find this effect to be significant compared to the effect of the inclusion of the thermal boundary resistance (see SI for details). Another possibility for additional heating in electronic components operating at cryogenic temperature is the relative scarcity of low-energy phonons necessary for heat dissipation.[38] As a result, local temperature is elevated until high-energy modes are populated and can dissipate heat at the required rate. Our simulations predict that the temperature increase of the sapphire substrate immediately adjacent to the nanowire is below ~0.5K (see SI for details). Inclusion of the self-heating effect in the model would therefore constitute a second order correction, difficult to discern at present as this temperature increase is comparable to the experimental precision and device-to-device variation.

**Conclusion**

We demonstrate a significant increase in the local temperature of plasmonically active Au nanowires under resonant laser radiation at low temperatures, with a nontrivial dependence on substrate temperature and incident laser intensity and polarization. Using full numerical modeling of the optical heat generation and thermal transport in these structures, we verify that the non-linearity in the laser intensity dependence of the temperature change is due to the non-linear temperature dependence of the relevant material properties. Our model and experimental results confirm that the thermal boundary resistance at the metal-substrate interface is a key parameter that limits thermal transport out of the metal nanostructure and determines the temperature increase at low temperature and low laser intensity. We stress the importance of the numerical modeling, as it highlights that the finite resistance of the leads connecting the nanowires introduces a systematic error in the bolometric detection scheme that either underestimates or overestimates the actual local temperature. We report a threefold reduction in optically driven temperature elevation by switching to the sapphire substrate from thermal oxide covered silicon wafers. This reduction is important in light of the fact that substantial temperature increase could lead to an unwanted thermal expansion in nanowires containing nanoscale gaps and undermine future single molecule transport experiments under laser radiation that require a cryogenic environment.

**Methods**

Experiments were performed in a home-built scanning Raman microscope equipped with an optical cryostat (Montana Instruments).[33,49] The devices were kept under high vacuum (<$10^{-7}$ bar) during the measurement. Polarization of laser radiation was rotated using a half-wave plate, with 0° polarization defined as when the electric field is oriented along the nanowire (pointing from one large electrode to the other). Each device was biased through the summing amplifier by a combination of small AC and DC voltages to measure electrical properties of the devices under laser illumination.

***Device fabrication.*** Sapphire (C-plane), quartz (Y-cut) (MTI Corp), and n-type Si wafers with a 200 nm thermally grown oxide layer (NIST) were used as substrates for bowtie devices. After shadow-mask evaporation of the Au/Ti contact pads with 45nm/5nm thickness, smaller features were fabricated using standard e-beam lithography techniques. Each chip contained 24 bowtie devices that share common ground. For insulating substrates an additional 6nm chromium conductive layer was thermally evaporated on top of the PSSA/PMMA layer to reduce charging during e-beam lithography. A 9% solution of poly(4-styrenesulfonic acid) (PSSA) in water was used as a spacer between the chromium and PMMA. The chromium layer was removed after e-beam writing by dissolving PSSA in water. After development the samples were oxygen plasma cleaned in the reactive ion etcher for 5 sec to remove PMMA residue. The final deposition of 14nm/1nm Au/Ti layer was performed using e-beam evaporation. Samples were mounted on the chip carrier using Apiezon N thermal grease and wire bonded using 25 μm gold wires.

***Simulation model.*** A Gaussian beam was set as electromagnetic source to interact with the metallic junction placed on either $SiO_2$/Si or sapphire substrates. The bowtie geometry was reproduced up to 3 μm from the center of the junction, a size limitation large enough to avoid finite simulation-space effects while remaining relatively quick to compute. The four-fold symmetry of the geometry was leveraged for computational efficiency and calculations were performed only over one quadrant of the structure. The simulation domain was truncated by employing Perfectly Matched Layers and Infinite Element Domain for the heat transfer module. The results of the simulation are not affected by the reduction of the simulation geometry, since the beam size limits the electromagnetic interaction area. The final temperature pattern in the whole device was calculated using the electromagnetic dissipation (from Maxwell's equations) as the heating source distribution for heat transfer equations. The thermal conductivity for Au was calculated from the Wiedemann-Franz law and agrees well with the previously published data for a 16nm thick Au film.[50] The 1nm Ti adhesion layer is not included in the model (see SI for discussion of this and $\Delta T$ calculations with 1nm Ti layer).

**Supporting Information**.  The Supporting Information is available free of charge *via* the Internet at http://pubs.acs.org

**Acknowledgments**. P.Z. and D.N. acknowledge support from ARO award W911-NF-13-0476.  D. N. acknowledges support from the Robert A. Welch Foundation Grant C-1636. A.A. and P.N. acknowledge support from the Robert A. Welch Foundation under grant C-1222 and from the Air Force Office of Scientific Research under grant FA9550-15-1-0022.

**Competing financial interests**. The authors declare no competing financial interests.

**References:**

(1) Scarano, S.; Mascini, M.; Turner, A. P. F.; Minunni, M. Surface Plasmon Resonance Imaging for Affinity-Based Biosensors. *Biosens. Bioelectron.* **2010**, *25*, 957–966.
(2) Abbas, A.; Linman, M. J.; Cheng, Q. New Trends in Instrumental Design for Surface Plasmon Resonance-Based Biosensors. *Biosens. Bioelectron.* **2011**, *26*, 1815–1824.
(3) Liu, Z.; Steele, J. M.; Srituravanich, W.; Pikus, Y.; Sun, C.; Zhang, X. Focusing Surface Plasmons with a Plasmonic Lens. *Nano Lett.* **2005**, *5*, 1726–1729.
(4) Maier, S. A.; Atwater, H. A. Plasmonics: Localization and Guiding of Electromagnetic Energy in Metal/dielectric Structures. *J. Appl. Phys.* **2005**, *98*.
(5) Hess, O.; Pendry, J. B.; Maier, S. A.; Oulton, R. F.; Hamm, J. M.; Tsakmakidis, K. L. Active Nanoplasmonic Metamaterials. *Nat. Mater.* **2012**, *11*, 573–584.

(6) Wei, H.; Xu, H. Hot Spots in Different Metal Nanostructures for Plasmon-Enhanced Raman Spectroscopy. *Nanoscale* **2013**, *5*, 10794–10805.
(7) Kumar, G. V. P. Plasmonic Nano-Architectures for Surface Enhanced Raman Scattering: A Review. *J. Nanophotonics* **2012**, *6*.
(8) Baffou, G.; Quidant, R. Thermo-Plasmonics: Using Metallic Nanostructures as Nano-Sources of Heat. *Laser Photonics Rev.* **2013**, *7*, 171–187.
(9) Bayazitoglu, Y.; Kheradmand, S.; Tullius, T. K. An Overview of Nanoparticle Assisted Laser Therapy. *Int. J. Heat Mass Transf.* **2013**, *67*, 469–486.
(10) Qiu, J.; Wei, W. D. Surface Plasmon-Mediated Photothermal Chemistry. *J. Phys. Chem. C* **2014**, *118*, 20735–20749.
(11) Donner, J. S.; Morales-Dalmau, J.; Alda, I.; Marty, R.; Quidant, R. Fast and Transparent Adaptive Lens Based on Plasmonic Heating. *ACS Photonics* **2015**, *2*, 355–360.
(12) Zhu, M.; Baffou, G.; Meyerbröker, N.; Polleux, J. Micropatterning Thermoplasmonic Gold Nanoarrays to Manipulate Cell Adhesion. *ACS Nano* **2012**, *6*, 7227–7233.
(13) Steinbrück, A.; Choi, J.-W.; Fasold, S.; Menzel, C.; Sergeyev, A.; Pertsch, T.; Grange, R. Plasmonic Heating with near Infrared Resonance Nanodot Arrays for Multiplexing Optofluidic Applications. *RSC Adv.* **2014**, *4*, 61898–61906.
(14) Kaya, S.; Weeber, J.-C.; Zacharatos, F.; Hassan, K.; Bernardin, T.; Cluzel, B.; Fatome, J.; Finot, C. Photo-Thermal Modulation of Surface Plasmon Polariton Propagation at Telecommunication Wavelengths. *Opt. Express* **2013**, *21*, 22269–22284.
(15) Sheldon, M. T.; Groep, J. van de; Brown, A. M.; Polman, A.; Atwater, H. A. Plasmoelectric Potentials in Metal Nanostructures. *Science* **2014**, *346*, 828–831.
(16) Bouillard, J.-S. G.; Dickson, W.; O'Connor, D. P.; Wurtz, G. A.; Zayats, A. V. Low-Temperature Plasmonics of Metallic Nanostructures. *Nano Lett.* **2012**, *12*, 1561–1565.
(17) Tian, J.-H.; Liu, B.; Li; Yang, Z.-L.; Ren, B.; Wu, S.-T.; Tao; Tian, Z.-Q. Study of Molecular Junctions with a Combined Surface-Enhanced Raman and Mechanically Controllable Break Junction Method. *J. Am. Chem. Soc.* **2006**, *128*, 14748–14749.
(18) Ioffe, Z.; Shamai, T.; Ophir, A.; Noy, G.; Yutsis, I.; Kfir, K.; Cheshnovsky, O.; Selzer, Y. Detection of Heating in Current-Carrying Molecular Junctions by Raman Scattering. *Nat. Nanotechnol.* **2008**, *3*, 727–732.
(19) Ward, D. R.; Halas, N. J.; Ciszek, J. W.; Tour, J. M.; Wu, Y.; Nordlander, P.; Natelson, D. Simultaneous Measurements of Electronic Conduction and Raman Response in Molecular Junctions. *Nano Lett.* **2008**, *8*, 919–924.
(20) Matsuhita, R.; Horikawa, M.; Naitoh, Y.; Nakamura, H.; Kiguchi, M. Conductance and SERS Measurement of Benzenedithiol Molecules Bridging Between Au Electrodes. *J. Phys. Chem. C* **2013**, *117*, 1791–1795.
(21) Klingsporn, J. M.; Jiang, N.; Pozzi, E. A.; Sonntag, M. D.; Chulhai, D.; Seideman, T.; Jensen, L.; Hersam, M. C.; Duyne, R. P. V. Intramolecular Insight into Adsorbate–Substrate Interactions *via* Low-Temperature, Ultrahigh-Vacuum Tip-Enhanced Raman Spectroscopy. *J. Am. Chem. Soc.* **2014**, *136*, 3881–3887.
(22) Wu, S. W.; Ogawa, N.; Ho, W. Atomic-Scale Coupling of Photons to Single-Molecule Junctions. *Science* **2006**, *312*, 1362–1365.
(23) Lee, J.; Perdue, S. M.; Whitmore, D.; Apkarian, V. A. Laser-Induced Scanning Tunneling Microscopy: Linear Excitation of the Junction Plasmon. *J. Chem. Phys.* **2010**, *133*, 104706.

(24) Carlson, M. T.; Khan, A.; Richardson, H. H. Local Temperature Determination of Optically Excited Nanoparticles and Nanodots. *Nano Lett.* **2011**, *11*, 1061–1069.
(25) Baffou, G.; Kreuzer, M. P.; Kulzer, F.; Quidant, R. Temperature Mapping near Plasmonic Nanostructures Using Fluorescence Polarization Anisotropy. *Opt. Express* **2009**, *17*, 3291–3298.
(26) Setoura, K.; Werner, D.; Hashimoto, S. Optical Scattering Spectral Thermometry and Refractometry of a Single Gold Nanoparticle under CW Laser Excitation. *J. Phys. Chem. C* **2012**, *116*, 15458–15466.
(27) Pozzi, E. A.; Zrimsek, A. B.; Lethiec, C. M.; Schatz, G. C.; Hersam, M. C.; Van, D. Evaluating Single-Molecule Stokes and Anti-Stokes SERS for Nanoscale Thermometry. *J. Phys. Chem. C* **2015**, *119*, 21116–21124.
(28) Coppens, Z. J.; Li, W.; Walker, D. G.; Valentine, J. G. Probing and Controlling Photothermal Heat Generation in Plasmonic Nanostructures. *Nano Lett.* **2013**, *13*, 1023–1028.
(29) Schmid, S.; Wu, K.; Larsen, P. E.; Rindzevicius, T.; Boisen, A. Low-Power Photothermal Probing of Single Plasmonic Nanostructures with Nanomechanical String Resonators. *Nano Lett.* **2014**, *14*, 2318–2321.
(30) Virk, M.; Xiong, K.; Svedendahl, M.; Käll, M.; Dahlin, A. B. A Thermal Plasmonic Sensor Platform: Resistive Heating of Nanohole Arrays. *Nano Lett.* **2014**, *14*, 3544–3549.
(31) Viarbitskaya, S.; Cuche, A.; Teulle, A.; Sharma, J.; Girard, C.; Arbouet, A.; Dujardin, E. Plasmonic Hot Printing in Gold Nanoprisms. *ACS Photonics* **2015**, *2*, 744–751.
(32) Baffou, G.; Girard, C.; Quidant, R. Mapping Heat Origin in Plasmonic Structures. *Phys. Rev. Lett.* **2010**, *104*, 136805.
(33) Herzog, J. B.; Knight, M. W.; Natelson, D. Thermoplasmonics: Quantifying Plasmonic Heating in Single Nanowires. *Nano Lett.* **2014**, *14*, 499–503.
(34) Swartz, E. T.; Pohl, R. O. Thermal Boundary Resistance. *Rev. Mod. Phys.* **1989**, *61*, 605–668.
(35) Swartz, E. T.; Pohl, R. O. Thermal Resistance at Interfaces. *Appl. Phys. Lett.* **1987**, *51*, 2200–2202.
(36) Chen, G. Nonlocal and Nonequilibrium Heat Conduction in the Vicinity of Nanoparticles. *J. Heat Transf.* **1996**, *118*, 539–545.
(37) Hu, Y.; Zeng, L.; Minnich, A. J.; Dresselhaus, M. S.; Chen, G. Spectral Mapping of Thermal Conductivity through Nanoscale Ballistic Transport. *Nat. Nanotechnol.* **2015**, *10*, 701–706.
(38) Schleeh, J.; Mateos, J.; Íñiguez-de-la-Torre, I.; Wadefalk, N.; Nilsson, P. A.; Grahn, J.; Minnich, A. J. Phonon Black-Body Radiation Limit for Heat Dissipation in Electronics. *Nat. Mater.* **2015**, *14*, 187–192.
(39) Ganser, A.; Benner, D.; Waitz, R.; Boneberg, J.; Scheer, E.; Leiderer, P. Time-Resolved Optical Measurement of Thermal Transport by Surface Plasmon Polaritons in Thin Metal Stripes. *Appl. Phys. Lett.* **2014**, *105*, 191119.
(40) Benner, D.; Boneberg, J.; Nürnberger, P.; Waitz, R.; Leiderer, P.; Scheer, E. Lateral and Temporal Dependence of the Transport through an Atomic Gold Contact under Light Irradiation: Signature of Propagating Surface Plasmon Polaritons. *Nano Lett.* **2014**, *14*, 5218–5223.
(41) Artur, C. G.; Miller, R.; Meyer, M.; Le Ru, E. C.; Etchegoin, P. G. Single-Molecule SERS Detection of C60. *Phys. Chem. Chem. Phys. PCCP* **2012**, *14*, 3219–3225.

(42) Banik, M.; El-Khoury, P. Z.; Nag, A.; Rodriguez-Perez, A.; Guarrottxena, N.; Bazan, G. C.; Apkarian, V. A. Surface-Enhanced Raman Trajectories on a Nano-Dumbbell: Transition from Field to Charge Transfer Plasmons as the Spheres Fuse. *ACS Nano* **2012**, *6*, 10343–10354.
(43) Li, Y.; Doak, P.; Kronik, L.; Neaton, J. B.; Natelson, D. Voltage Tuning of Vibrational Mode Energies in Single-Molecule Junctions. *Proc. Natl. Acad. Sci. U. S. A.* **2014**, *111*, 1282–1287.
(44) Berto, P.; Mohamed, M. S. A.; Rigneault, H.; Baffou, G. Time-Harmonic Optical Heating of Plasmonic Nanoparticles. *Phys. Rev. B* **2014**, *90*, 035439.
(45) Alabastri, A.; Tuccio, S.; Giugni, A.; Toma, A.; Liberale, C.; Das, G.; Angelis, F. D.; Fabrizio, E. D.; Zaccaria, R. P. Molding of Plasmonic Resonances in Metallic Nanostructures: Dependence of the Non-Linear Electric Permittivity on System Size and Temperature. *Materials* **2013**, *6*, 4879–4910.
(46) Alabastri, A.; Toma, A.; Malerba, M.; De, A.; Proietti, Z. High Temperature Nanoplasmonics: The Key Role of Nonlinear Effects. *ACS Photonics* **2015**, *2*, 115–120.
(47) Hopkins, P. E.; Phinney, L. M.; Serrano, J. R.; Beechem, T. E. Effects of Surface Roughness and Oxide Layer on the Thermal Boundary Conductance at Aluminum/silicon Interfaces. *Phys. Rev. B - Condens. Matter Mater. Phys.* **2010**, *82*.
(48) Xu, Y.; Kato, R.; Goto, M. Effect of Microstructure on Au/sapphire Interfacial Thermal Resistance. *J. Appl. Phys.* **2010**, *108*.
(49) Evans, K. M.; Zolotavin, P.; Natelson, D. Plasmon-Assisted Photoresponse in Ge-Coated Bowtie Nanojunctions. *ACS Photonics* **2015**, *2*, 1192–1198.
(50) Zink, B. L.; Revaz, B.; Cherry, J. J.; Hellman, F. Measurement of Thermal Conductivity of Thin Films with a Si-N Membrane-Based Microcalorimeter. *Rev. Sci. Instrum.* **2005**, *76*, 024901.

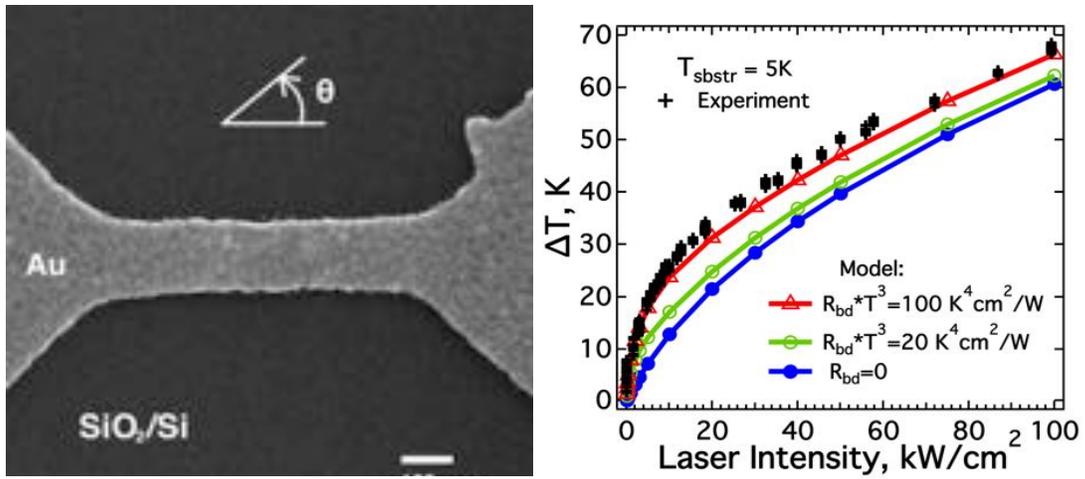

TOC Figure

# Supporting Information

# Plasmonic Heating in Au Nanowires at Low Temperatures: The Role of Thermal Boundary Resistance

Pavlo Zolotavin[1], Alessandro Alabastri[1], Peter Nordlander[1,2,3], Douglas Natelson[*,1,2,3]

[1]Department of Physics and Astronomy, Rice University, 6100 Main St., Houston, Texas 77005, United States

[2]Department of Electrical and Computer Engineering, Rice University, 6100 Main St., Houston, Texas 77005, United States

[3]Department of Materials Science and NanoEngineering, Rice University, 6100 Main St., Houston, Texas 77005, United States

*To whom the correspondence should be addressed. E-mail: natelson@rice.edu.

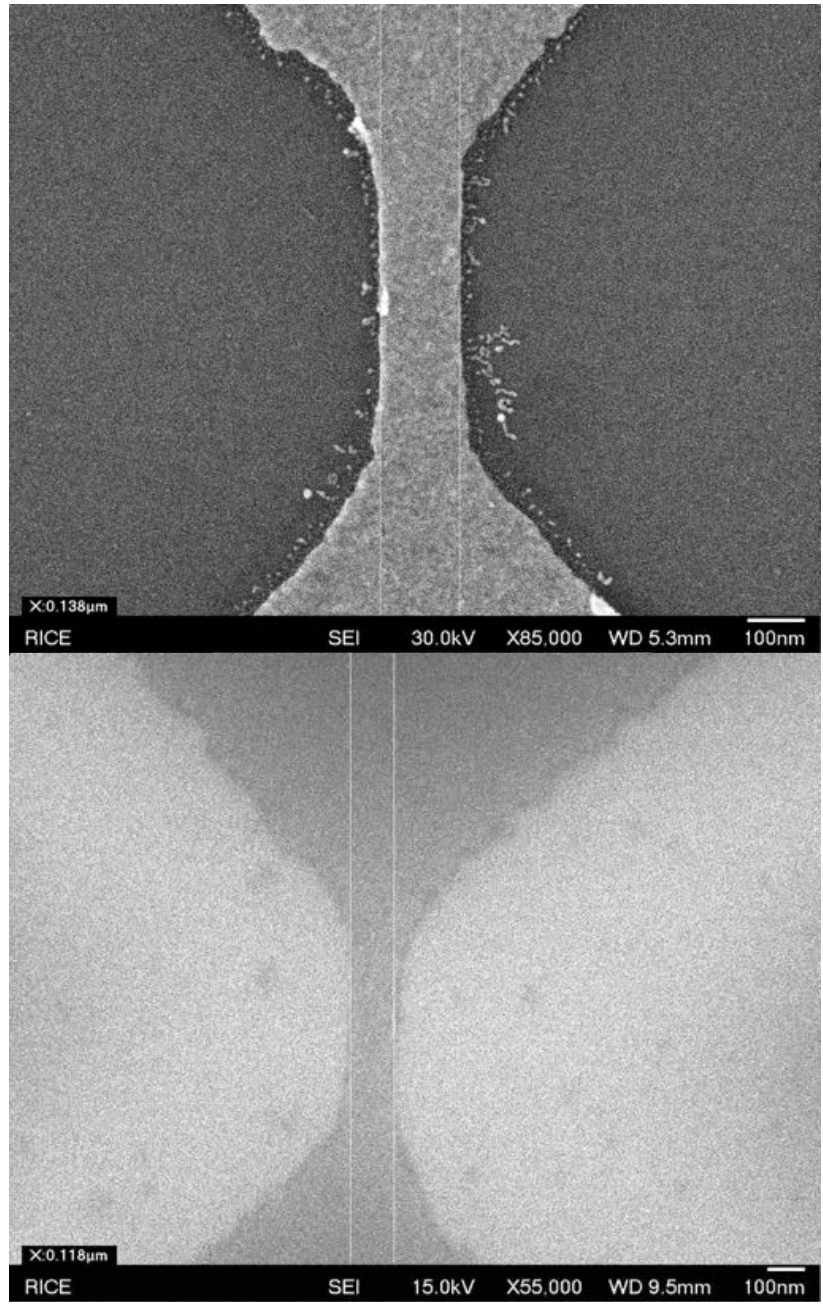

**Fig. S1**. SEM images of devices fabricated on quartz (top) and sapphire substrates (bottom). The device width is 138 ± 5 nm and 120 ± 5 nm respectively. The images correspond to the devices for which the data is presented in main text of the paper and here.

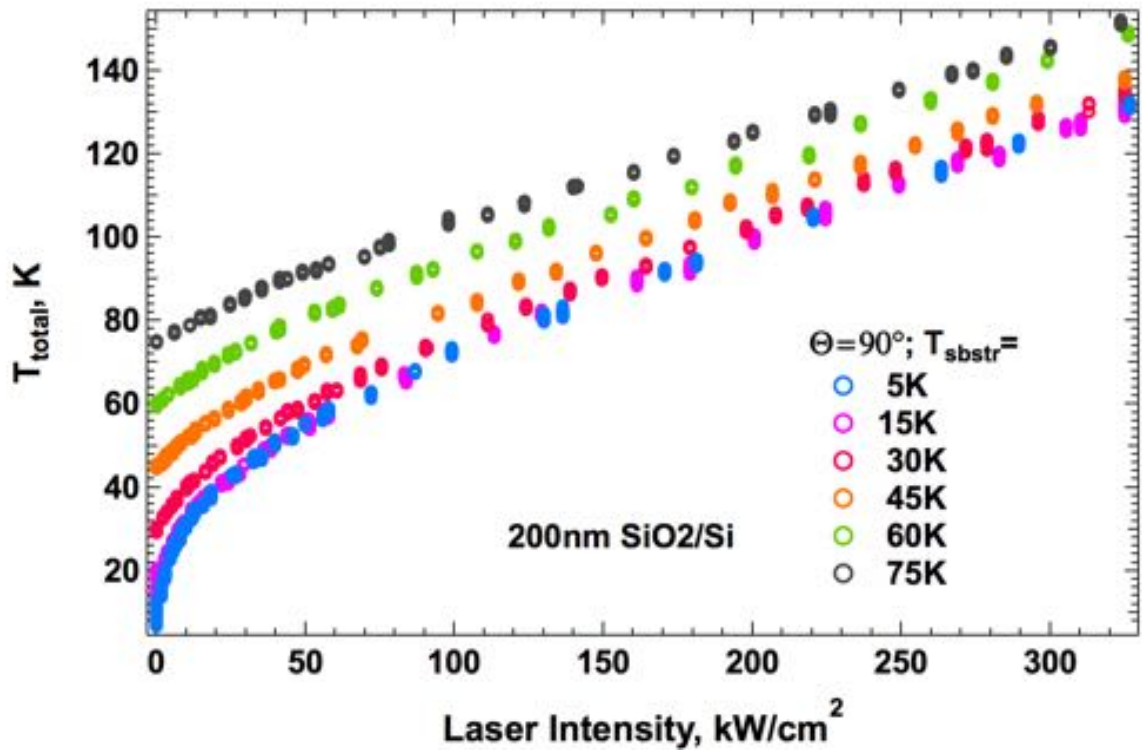

**Fig. S2** Dependence of the total temperature of the device on laser intensity for $SiO_2/Si$ substrate. Data is replotted from Fig 1c in the main text.

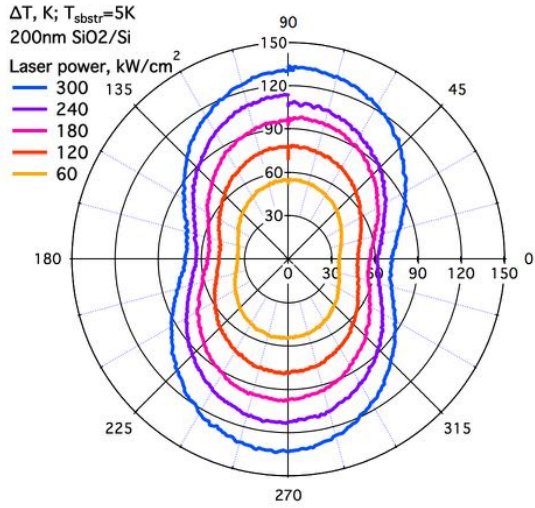
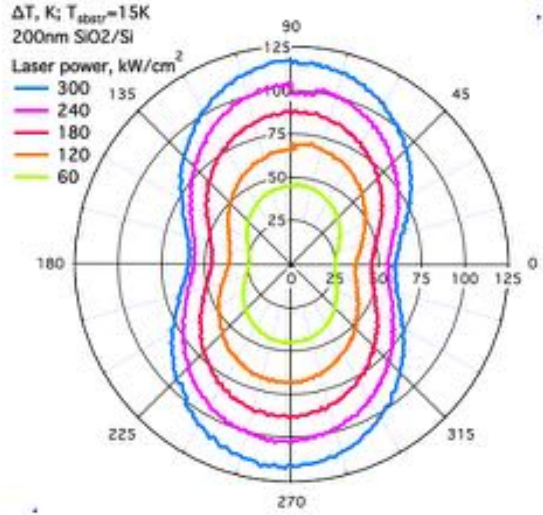
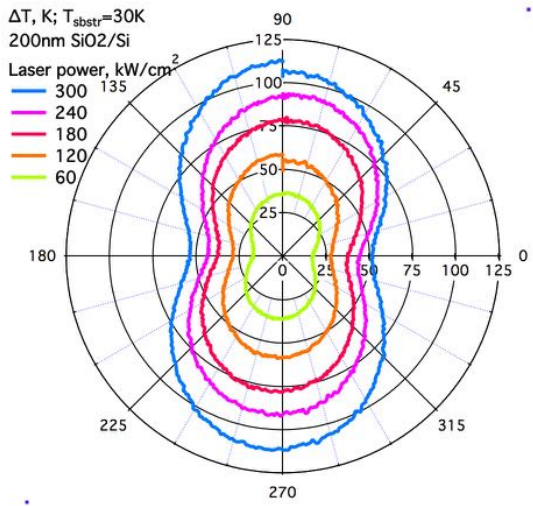
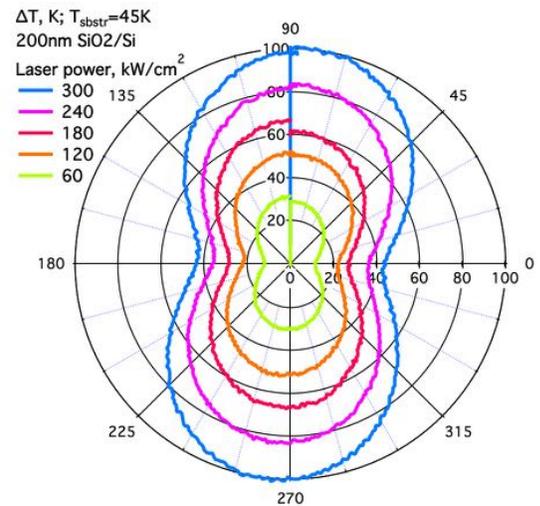
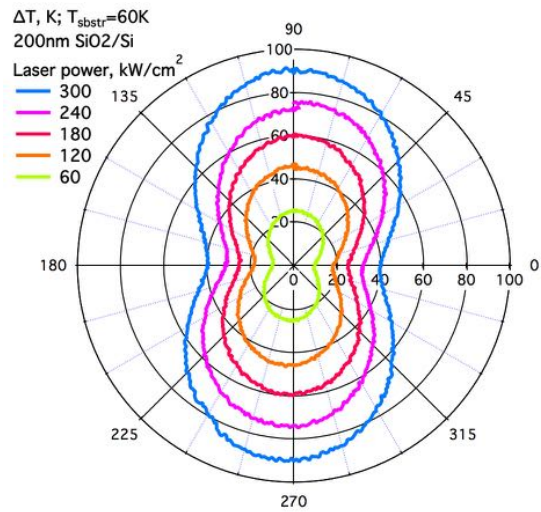

**Fig. S3**. Polarization dependence of the inferred average nanowire temperature increase for devices fabricated on $SiO_2$/Si substrate at different substrate temperatures (5–60 K) and laser intensity (60–300 kW/cm$^2$).

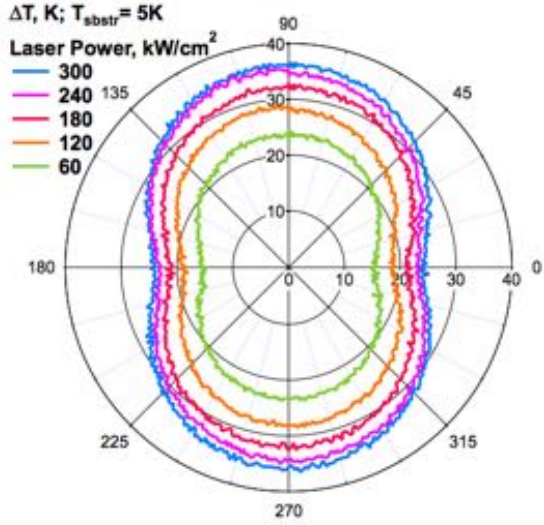
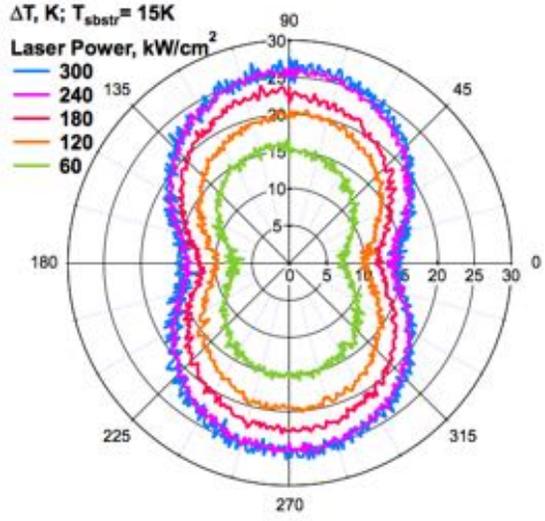
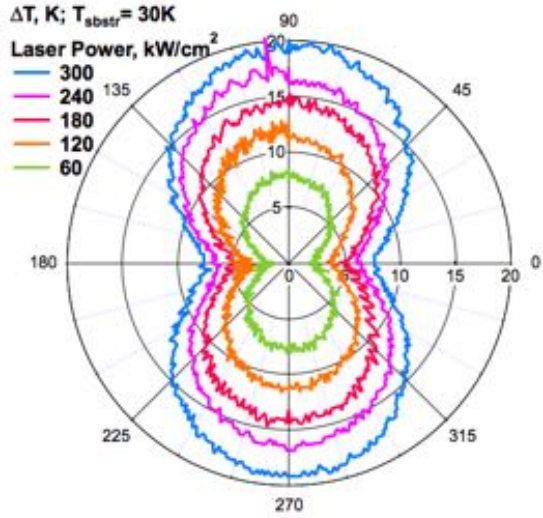
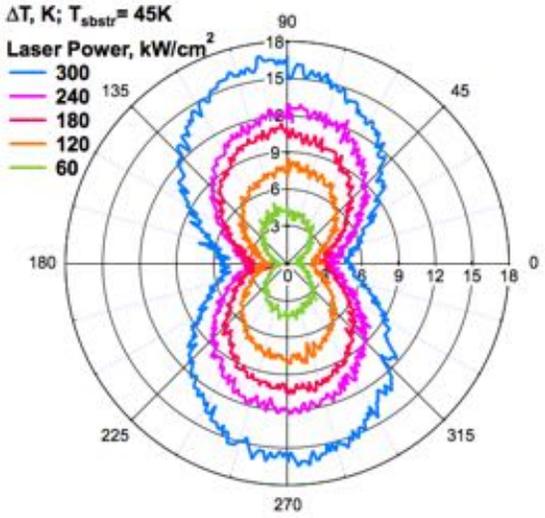
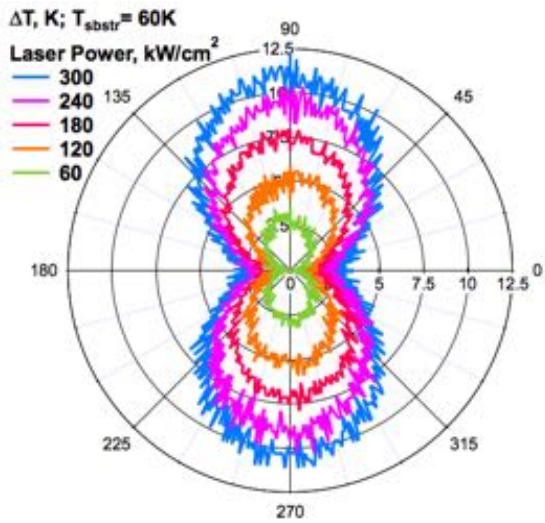

**Fig. S4**. Polarization dependence of the inferred average nanowire temperature increase for devices fabricated on sapphire substrate at different substrate temperatures (5–60 K) and laser intensity (60–300 kW/cm$^2$).

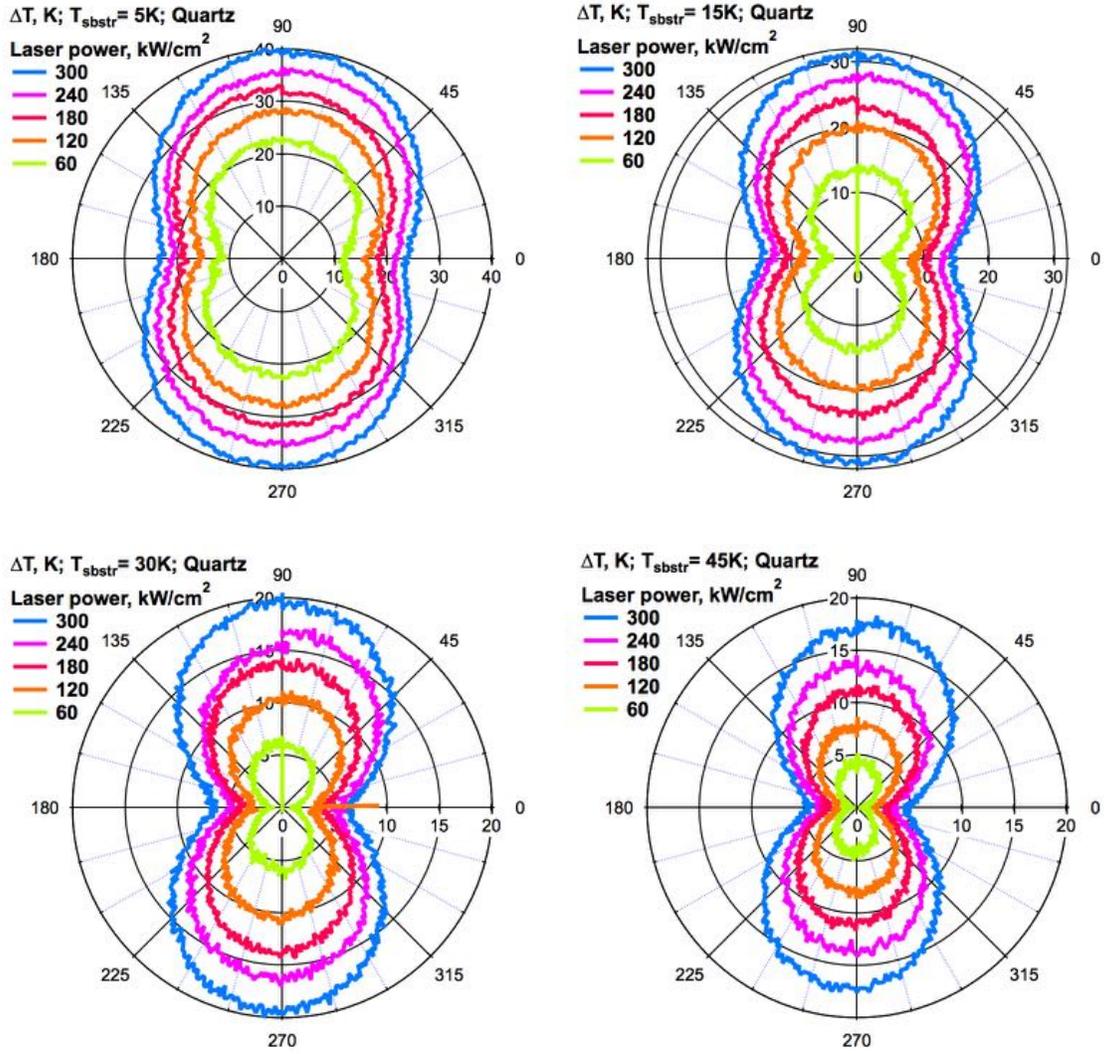

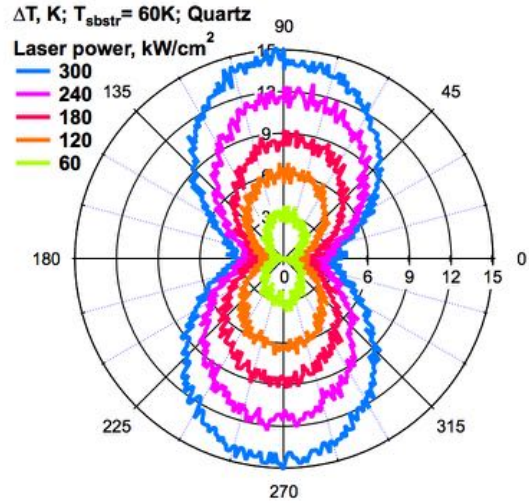

**Fig. S5**. Polarization dependence of the inferred average nanowire temperature increase for devices fabricated on quartz substrate at different substrate temperatures (5–60 K) and laser intensity (60–300 kW/cm²).

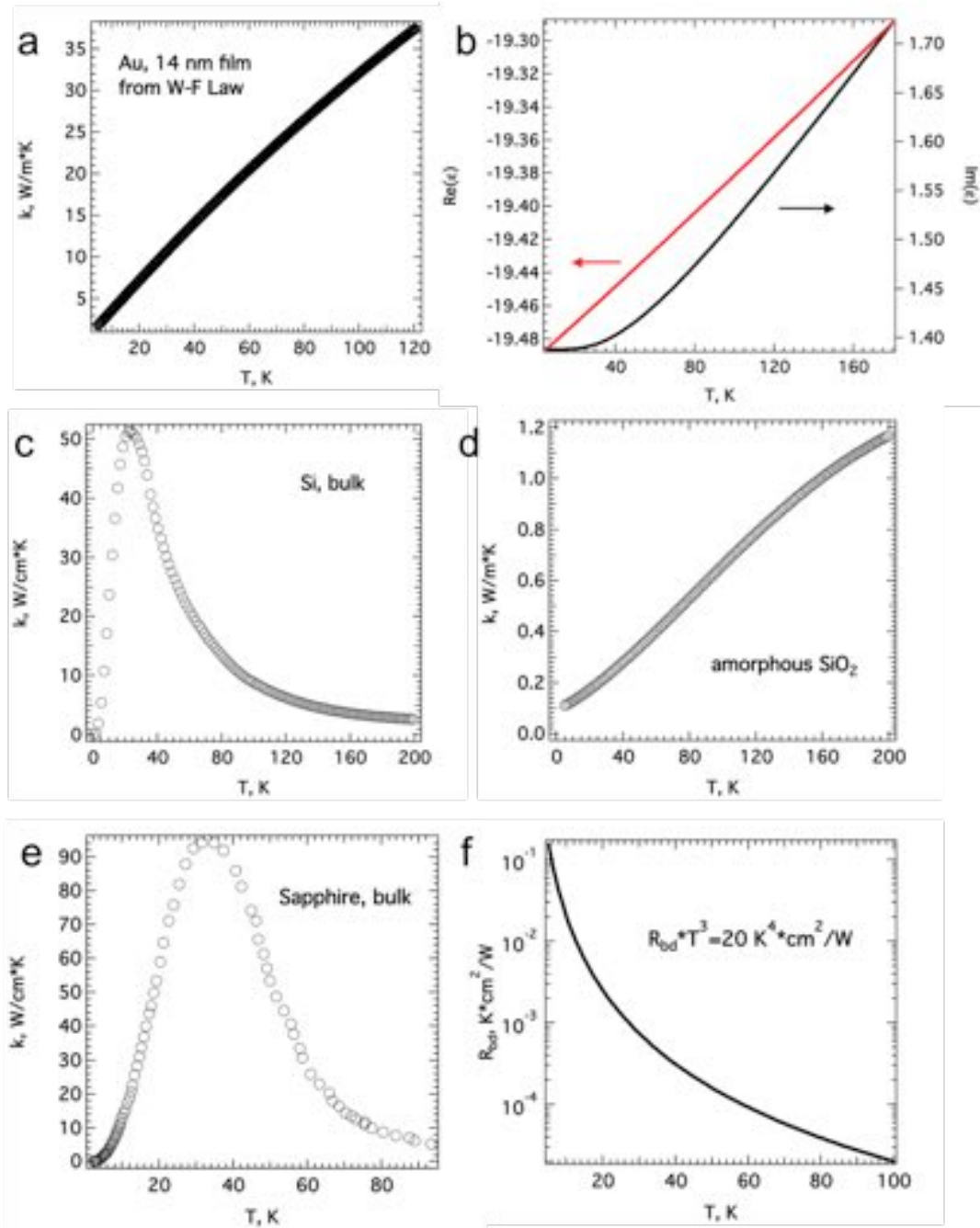

**Fig. S6.** Temperature dependent material properties that were used in the model: (a) thermal conductivity of 14nm gold film calculated using a Wiedemann-Franz law, (b) real and imaginary part of the dielectric constant of gold, calculated following the method outlined in Ref. 1, (c) thermal conductivity of bulk Si,[2] (d) thermal conductivity of amorphous $SiO_2$,[3,4] (e) thermal conductivity of bulk sapphire crystal,[5] (f) theoretical prediction of the thermal boundary resistance based on the acoustic mismatch model.[6]

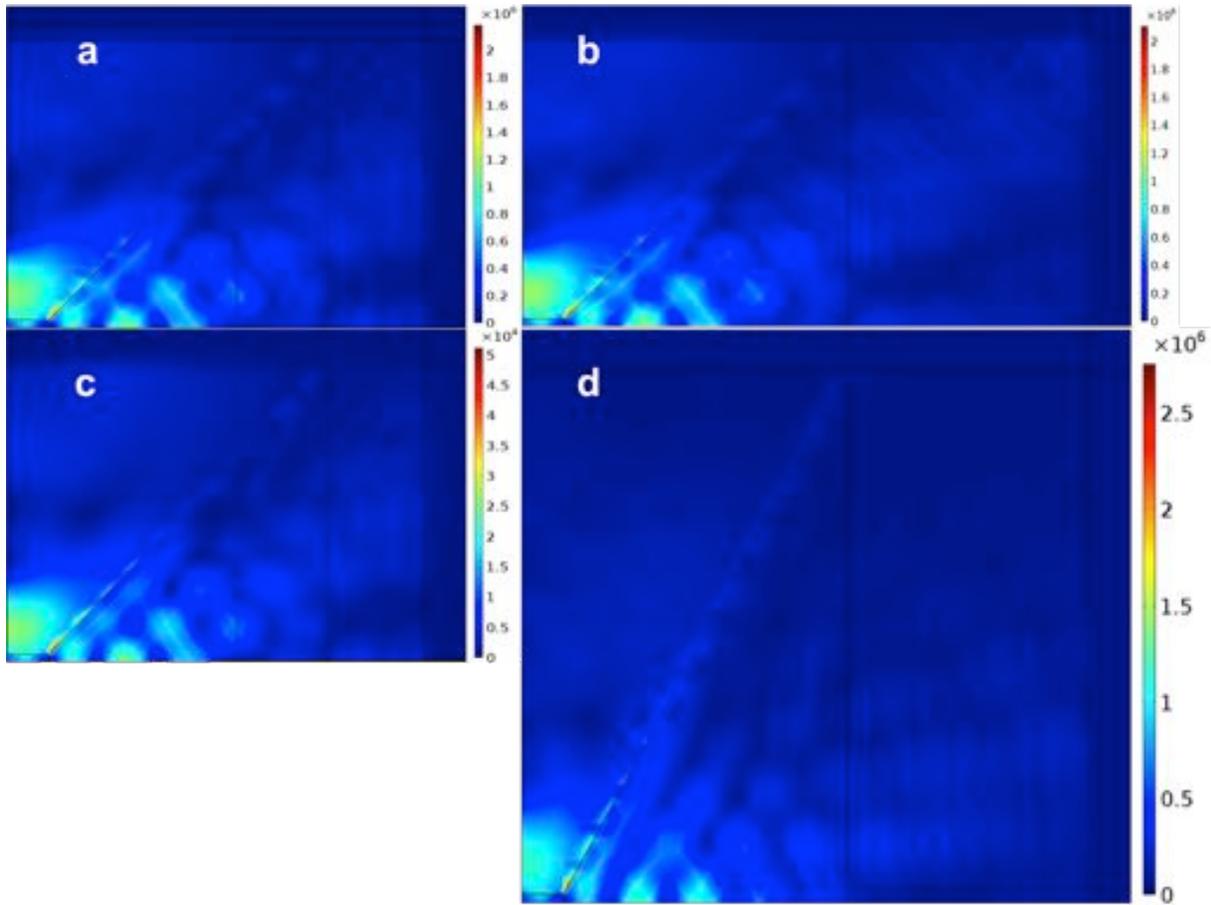

**Fig. S7**. Robustness of the observed oscillations in the $\Delta T(d)$ against changes in simulation geometry: (a) Intensity distribution of the electric field (V/m) for data displayed in the Fig. 3b of the main text; (b) the same, but with a longer connecting pad; (c) the same as in (a), but with a finer mesh; (d) the same as in (b), but with a wider connection pad.

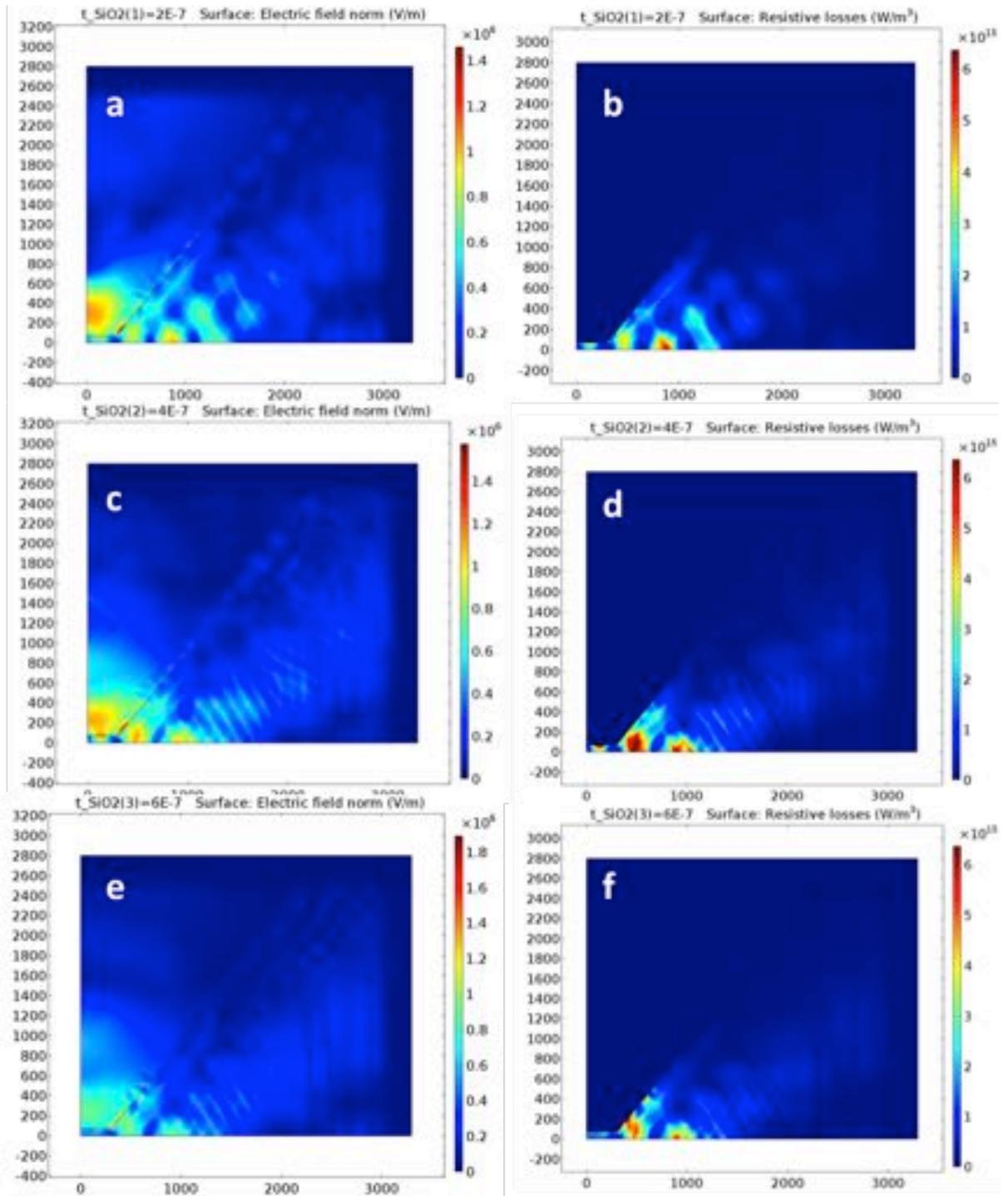

**Fig. S8**. Effect of the different SiO$_2$ thickness on the observed oscillations in the $\Delta T(d)$: (a, b) Intensity distribution of the electric field and surface resistive losses for data displayed

in the Fig. 3b of the main text, SiO$_2$ thickness is 200 nm; (c, d) the same, but with 400 nm SiO$_2$ layer thickness; (e, f) the same, but with 600 nm SiO$_2$ layer thickness;

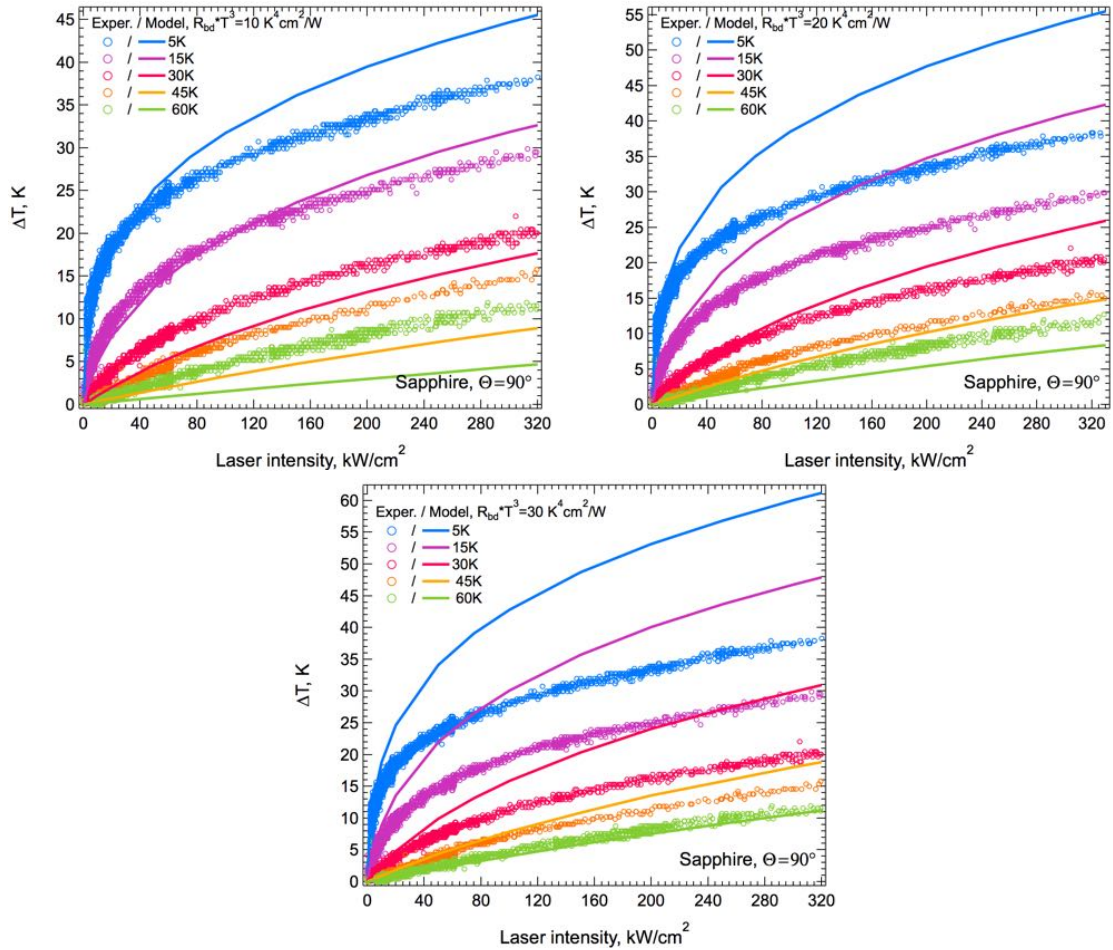

**Fig. S9**. Comparison of the experimental data for sapphire substrate laser intensity dependence with calculation results for $R_{bd}T^3$ in the range from 10 to 30 $K^4cm^2/W$ and at different substrate temperatures (5–60 K). Recall that the temperature inferred from the bolometric approach for polarization of 90° underestimates the true local temperature.

The choice of the exact value for the $R_{bd}$ significantly influences the predicted temperature rice due to laser illumination. The acoustic mismatch model sets the lower boundary for the value of $R_{bd}$. A large volume of experimental evidence suggests that for solid-solid interfaces $R_{bd}$ is usually larger than predicted by theory.[6] The value for the upper boundary we can deduce from the following argument. The cubic temperature dependence as predicted by the acoustic mismatch model was demonstrated to be a good approximation only up to $T \sim 30\,K$.[7] Above this temperature the thermal boundary resistance is larger than predicted by a cubic temperature dependence, which should lead to calculations that underestimate the temperature increase at $T_{sbstr} > 45$ K. This is already evident for $R_{bd}T^3 = 30\ K^4 cm^2/W$. We, therefore, conclude that the $R_{bd}T^3$ in our experimental system is within the 19 and 30 $K^4 cm^2/W$. This uncertainty leads to a ~5K error in the calculated temperature increase, which given the magnitude of $\Delta T$ constitutes only a ~10% error.

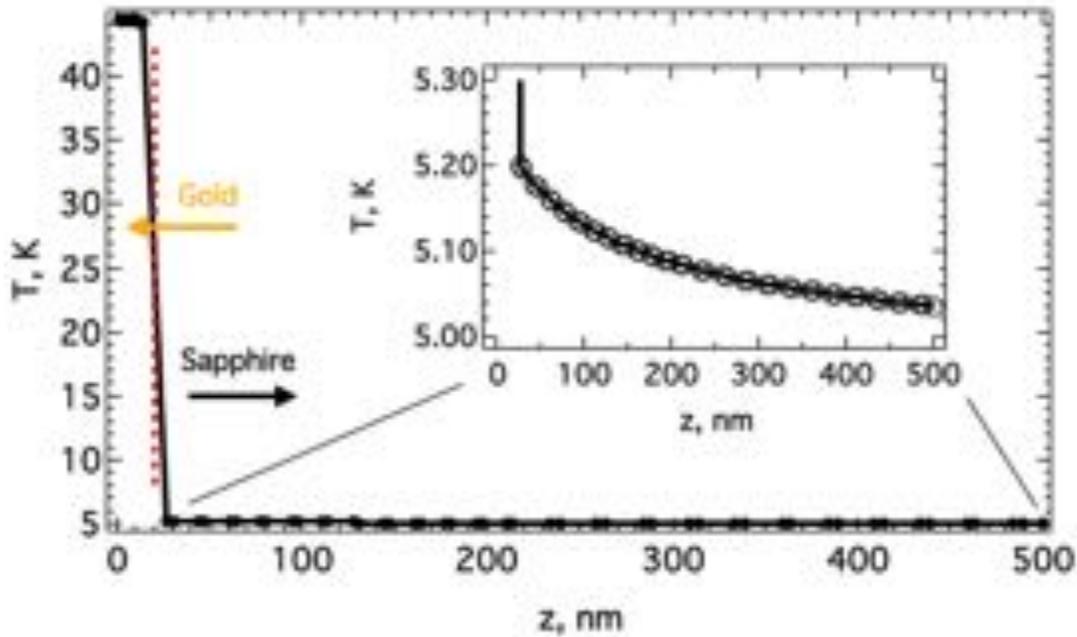

**Fig. S10**. Temperature variation along the vertical line going through the center of the nanowire into sapphire substrate. Inset is a zoom in on the bottom section of the figure. The data corresponds to the laser intensity of 100 $kW/cm^2$ and $T_{sbstr}$ = 5K demonstrated in Fig 5b in the main text.

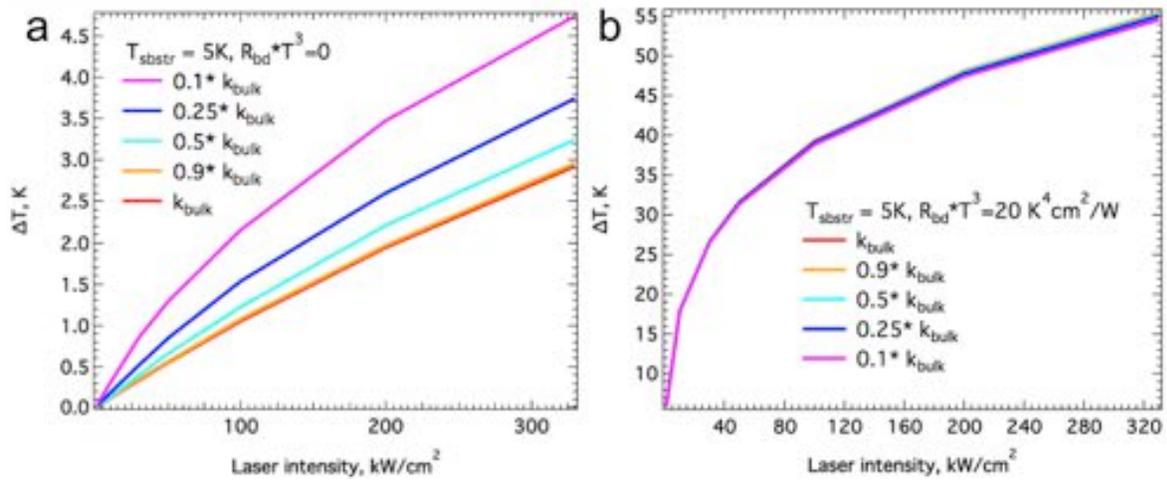

**Fig. S11**. Demonstration of the minimal effect of the reduction of the bulk of thermal conductivity of sapphire substrate on temperature increase for a model without, (a), and with, (b), thermal boundary resistance.

**Remarks regarding the omission of the 1 nm Ti adhesion layer from the model**

We chose not to include the adhesion Ti layer in the model based on the following:

- An average thickness of 1nm amounts to 2-4 atomic layers, and the film is likely to be discontinuous at this coverage. The first 1-2 of which are going to be in the form of $Ti_xO_y$ from binding to the oxygen in the substrate and the top 1-2 layers will be alloyed with gold. This leaves a layer of 0-2 atoms thick with poorly defined optical and thermal properties.
- It was demonstrated that addition of the adhesion metal to the gold-sapphire interface could reduce the thermal boundary resistance at room temperature.[8] We do not observe this effect, which serves as an incidental evidence to support that we don't have a properly formed continuous layer of Ti with well-defined bulk properties.
- An estimate of the device temperature based on the ratio between the anti-Stokes and Stokes Raman background level yields nanowire temperature of 130 ± 10K at maximum incident laser intensity. This number agrees well with model without Ti layer.

Below we demonstrate the effects of the inclusion of the additional 1nm Ti layer using the room temperature bulk material properties. The latter serves as an upper boundary for possible additional dissipation.

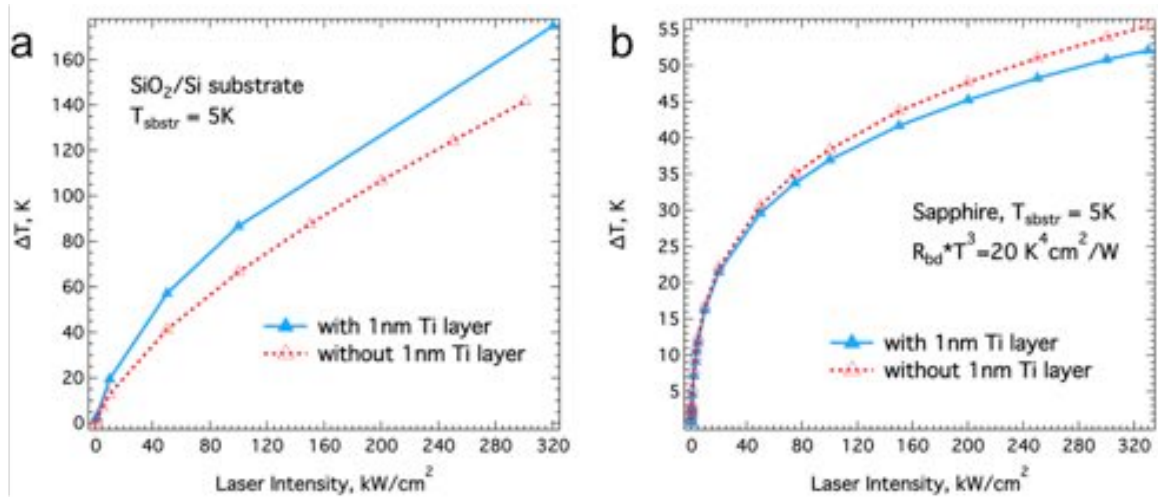

**Fig. S11** Change in the predicted magnitude of heating after inclusion in the model of the 1nm Ti adhesion layer for SiO$_2$/Si, (a), and sapphire, (b), substrates. Calculation for SiO$_2$/Si substrate is performed using full coupling model.

**References:**
(1) Bouillard, J.-S. G.; Dickson, W.; O'Connor, D. P.; Wurtz, G. A.; Zayats, A. V. Low-Temperature Plasmonics of Metallic Nanostructures. *Nano Lett.* **2012**, *12*, 1561–1565.
(2) Ho, C. Y.; Powell, R. W.; Liley, P. E. Thermal Conductivity of the Elements. *J. Phys. Chem. Ref. Data* **1972**, *1*, 279–421.
(3) Childs, G. E.; Ericks, L. J.; Powell, R. L. *Thermal Conductivity of Solids at Room Temperature and below: A Review and Compilation of the Literature*; National Bureau of Standards.
(4) Lee, D. W.; Kingery, W. D. Radiation Energy Transfer and Thermal Conductivity of Ceramic Oxides. *J. Am. Ceram. Soc.* **1960**, *43*, 594–607.
(5) Berman, R.; Foster, E. L.; Ziman, J. M. Thermal Conduction in Artificial Sapphire Crystals at Low Temperatures. I. Nearly Perfect Crystals. *Proc. R. Soc. Lond. Math. Phys. Eng. Sci.* **1955**, *231*, 130–144.
(6) Swartz, E. T.; Pohl, R. O. Thermal Boundary Resistance. *Rev. Mod. Phys.* **1989**, *61*, 605–668.
(7) Swartz, E. T.; Pohl, R. O. Thermal Resistance at Interfaces. *Appl. Phys. Lett.* **1987**, *51*, 2200–2202.
(8) Jeong, M.; Freedman, J. P.; Liang, H. J.; Chow, C.-M.; Sokalski, V. M.; Bain, J. A.; Malen, J. A. Enhancement of Thermal Conductance at Metal-Dielectric Interfaces Using Subnanometer Metal Adhesion Layers. *Phys. Rev. Appl.* **2016**, *5*, 014009.